\documentclass[prd,superscriptaddress,showpacs,nofootinbib,amsmath,amssymb,aps,11pt]{revtex4}

\usepackage{bm}
\usepackage{amsfonts}
\usepackage{latexsym}
\usepackage[latin1]{inputenc}
\usepackage{graphicx}
\usepackage{amsmath}
\usepackage{palatino}
\usepackage{mathpazo}
\usepackage{booktabs}
\usepackage{dcolumn}
\usepackage[outdir=./]{epstopdf}

\def\jnl@style{\it}
\def\aaref@jnl#1{{\jnl@style#1}}

\def\aaref@jnl#1{{\jnl@style#1}}

\def\aj{\aaref@jnl{AJ}}                   
\def\apj{\aaref@jnl{ApJ}}                 
\def\apjl{\aaref@jnl{ApJ}}                
\def\apjs{\aaref@jnl{ApJS}}               
\def\apss{\aaref@jnl{Ap\&SS}}             
\def\aap{\aaref@jnl{A\&A}}                
\def\aapr{\aaref@jnl{A\&A~Rev.}}          
\def\aaps{\aaref@jnl{A\&AS}}              
\def\mnras{\aaref@jnl{Mon.~Not.~Roy.~Astron.~Soc.}}             
\def\prd{\aaref@jnl{Phys.~Rev.~D}}        
\def\prc{\aaref@jnl{Phys.~Rev.~C}}  
\def\prl{\aaref@jnl{Phys.~Rev.~Lett.}}    
\def\qjras{\aaref@jnl{QJRAS}}             
\def\skytel{\aaref@jnl{S\&T}}             
\def\ssr{\aaref@jnl{Space~Sci.~Rev.}}     
\def\zap{\aaref@jnl{ZAp}}                 
\def\nat{\aaref@jnl{Nature}}               
\def\aplett{\aaref@jnl{Astrophys.~Lett.}} 
\def\apspr{\aaref@jnl{Astrophys.~Space~Phys.~Res.}} 
\def\physrep{\aaref@jnl{Phys.~Rep.}}      
\def\physscr{\aaref@jnl{Phys.~Scr}}       
\def\commat{\aaref@jnl{Comm.~Math.~Phys.}}              
\def\science{\aaref@jnl{Science}}               
\def\cqg{\aaref@jnl{Classical Quant.~Grav.}}            
\def\jpcs{\aaref@jnl{JPCS}}                                     
\def\ijmpd{\aaref@jnl{Int.~J.~Mod.~Phys.~D}}                    
\def\grg{\aaref@jnl{Gen.~Relat.~Gravit.}}               
\def\rpp{\aaref@jnl{Rep.~Prog.~Phys.}}          
\def\npa{\aaref@jnl{Nucl.~Phys.~A}}        
\def\lrr{\aaref@jnl{Living Rev.~Rel.}}                   
\def\jcap{\aaref@jnl{J.~Cosmology Astropart.~Phys.}}    
\def\rmp{\aaref@jnl{Rev.~Mod.~Phys.}}   

\allowdisplaybreaks[1]

\begin{document}

\title{Asteroseismology of rapidly rotating neutron stars -- an alternative approach}

\author{Daniela D. Doneva}
\email{daniela.doneva@uni-tuebingen.de}
\affiliation{Theoretical Astrophysics, Eberhard Karls University of T\"ubingen, T\"ubingen 72076, Germany}
\affiliation{INRNE - Bulgarian Academy of Sciences, 1784  Sofia, Bulgaria}

\author{Kostas D. Kokkotas}
\affiliation{Theoretical Astrophysics, Eberhard Karls University of T\"ubingen, T\"ubingen 72076, Germany}


\begin{abstract}
In the present paper we examine gravitational wave asteroseismology relations for $f$-modes of rapidly rotating neutron stars. An approach different to the previous studies is employed -- first, the moment of inertia is used instead of the stellar radius, and second, the normalization of the oscillation frequencies and damping times is different. It was shown that in the non-rotating case this can lead to a much stronger equation of state independence and our goal is to generalize the static relations to the rapidly rotating case and values of the spherical mode number $l\ge2$. We employ realistic equations of state that cover a very large range of stiffness in order to check better the universality of the relations. At the end we explore the inverse problem, i.e. obtain the neutron star parameters from the observed gravitational frequencies and damping times. It turns out that with this new set of relations we can solve the inverse problem with a very good accuracy using three frequencies that was not possible in the previous studies where one needs also the damping times. The asteroseismology relations are also particularly good for the massive rapidly rotating models that are subject to secular instabilities.
\end{abstract}


\maketitle

\section{Introduction}
A direct detection of gravitational waves is a major goal in physics for several decades. Naturally the investigation of the strong sources of gravitational radiation is being done in parallel. Different dynamical processes connected to neutron stars, such as neutron star mergers and rotational instabilities, are considered as promising sources. The final goal is to use the observed gravitational wave signal in order to extract the characteristic parameters of the observed objects, such as mass, radius, rotational rate, etc., and eventually to constrain the nuclear matter equation of state (EoS) \cite{Bauswein2014,Takami2014,Andersson1996,Andersson98a,Kokkotas01,Gaertig10,Doneva2013a}. 

In the present paper we will concentrate on the neutron star oscillations with an emphasis on the region of the parameter space where the Chandrasekhar-Friedman-Schutz (CFS) instability can develop, i.e. certain nonaxisymmetric modes can become unstable due to the emission of gravitational waves. This is one of the most promising scenario for the emission of strong gravitational radiation where the signal can reach even above the Advance LIGO  sensitivity \cite{Passamonti12,DPK2015}. More specifically we will reexamine the gravitational wave asteroseismology with rapidly rotating neutron stars by considering a different parametrization of the EoS independent relations. 

Studying gravitational wave asteroseismology of oscillating neutron stars originates with the papers \cite{Andersson1996,Andersson98a,Kokkotas01} where the mode oscillation frequencies and damping times of static neutron stars were related to their mass and radius. It was found that the asteroseismology relations do not depend on the EoS up to a large extend and the inverse problem can be solved with a very good accuracy. Later these results were extended by including other modern realistic EoS \cite{Benhar04}. The generalization of the relations to the case of rapidly rotating neutron stars and larger values of the spherical mode number $l$ was done recently in \cite{Gaertig10,Doneva2013a}. These results extend up to the Kepler limit which means that asteroseismology can be performed for a much larger class of objects and more specifically for neutron stars that are subject to the CFS instability. In a similar fashion the frequencies of the emitted gravitational wave signal during binary neutron star mergers can be connected to the properties of the post-merger supramassive neutron stars \cite{Bauswein2014,Takami2014}.

All the relations obtained in \cite{Andersson98a,Benhar04,Gaertig10,Doneva2013a} connect the oscillation mode frequency and damping time to the neutron star mass, radius and rotational rate. But a different approach was proposed in \cite{Lau2010}  which steamed out of an earlier study by \citet{Lattimer2005} -- to  use the moment of inertia instead of the stellar radius. In this way the obtained relations are much more EoS insensitive compared to the previous results. Similar relations were also obtained in \cite{Chirenti2015}. But the results in \cite{Lau2010,Chirenti2015} are limited to the nonrotating case. Our main goal in this paper is to extend them to the rapidly rotating case in order to see if the EoS universality is preserved and whether this approach is more beneficial compared to the standard case \cite{Gaertig10,Doneva2013a}.

\section{Asteroseismology using the moment of inertia}
Our aim is to present the results as a brief report. For this reason we will not go into details about the method of calculation of the mode frequencies and damping times, but instead we will give only the most important points. We refer the reader to \cite{Doneva2013a} where a detail presentation of the methodology can be found (see also \cite{Gaertig10,Gaertig11,Kruger10}).

We obtain the oscillation modes by performing a time evolution of the linearized perturbation equations in the Cowling approximation, where the perturbations of the metric are neglected. The reason for these simplifications is that the full nonlinear simulations are extremely computer and manpower demanding \cite{Zink10} and solving the linearized perturbation equations without any additional approximation is not done until now. But we should note that even though this approximation can lead to relatively large deviations of the frequencies (in the range $10-30\%$ depending on the compactness and the mode number $l$) it was shown in \cite{Gaertig10,Doneva2013a} that it gives not only qualitatively but in some cases also quantitatively good results for the asteroseismology relations. 

The numerical code we use for the time evolution of the perturbation equations was developed in \cite{Kruger10,Doneva2013a}. The background neutron star solutions are obtained with the {\tt RNS} code \cite{Stergioulas97} that has been proven to be reliable for rapid rotation. The mode frequencies are easily obtained after performing a Fourier transform on the computed time series. Obtaining the damping time on the other hand is more involved. Since we are working in the Cowling approximation, where the background metric is not perturbed, the damping (or the growth in the case of CFS instability) of the modes due to the emission of gravitational waves can not be calculated directly \footnote{It is questionable whether this damping/growth time can be accurately calculated using a time evolution code in general, since the inevitable numerical dissipations in many cases are much stronger than the dissipation of energy due to gravitational wave emission.}. Instead one can apply approximate Newtonian formula, where the emission of gravitational waves is related to the mutlipole moments of the neutron star. It was shown that at least in the nonrotating case this formula gives good results for the damping time compared to the exact general relativistic values \cite{Balbinski85}. 

We use three EoS that cover a very large range of masses and radii -- EoS A \cite{A}, WFF2 \cite{WFF} and L \cite{Pandharipande1976}. The first one is very soft reaching a maximum mass of only 1.67 $M_\odot$. The second one is a standard modern realistic EoS that fulfills all the observational constraints on the neutron star mass and radius \cite{Lattimer12,Steiner2013,Antoniadis13,Demorest10}. EoS L on the other hand is a very stiff (one of the stiffest available in the literature) with $M_{max}=2.72 M_\odot$ and typical radii around $14-15$ km. The set of EoS is chosen to cover such large range of stiffness for a reason -- we want to verify up to what extend our results are insensitive to the EoS. For practical purposes one can of course use a narrower set of EoS that falls into the preferred range of masses and radii according to the observations. This can significantly reduce the error in the fitting formulae. Since we are calculating the modes in the Cowling approximation, we can not reduce the errors of our asteroseismology formulae below the deviations coming from this approximation. That is why we decided not to give separate fittings for a set of EoSs that fall into the preferred range of neutron star masses and radii. Our purpose here is more to offer an alternative way for doing asteroseismology of rapidly rotating neutron stars and to demonstrate its validity over a large range of EoS stiffness. 

Throughout the paper we will use the following dimensions for the different quantities. The mass $M$ is measured in solar masses $M_\odot$, the frequencies (the angular rotational frequency $\Omega$ as well as the mode oscillation frequency $\sigma$) are measured in kHz, the damping/growth time -- in second and the moment of inertia $I$ is defined as $I=I_{cgs}/10^{45}\,{\rm g\,cm^2}$, where $I_{cgs}$ is the moment of inertia in cgs units. In this way for example the quantity $\eta$, that is widely used in the paper and roughly speaking has a meaning of an effective compactness of the star, takes the form
\begin{equation}
\eta = \sqrt{\left(\frac{M}{M_\odot}\right)^3\left(\frac{I_{cgs}}{10^{45}{\rm g\,cm^2}}\right)^{-1}}.
\end{equation}
\subsection{Oscillation frequencies}
Here we will present the results for the asteroseismology relations of the oscillation frequencies. In the previous studies of asteroseismology with rapidly rotating neutron stars \cite{Gaertig10,Doneva2013a} two step relations were constructed that map the oscillation frequencies of rotating neutron stars to their mass, radius and rotational rate. As a first step we have relations between the normalized oscillation frequencies and the rotational rate derived for different values of $l$, where the oscillation frequencies are normalized to the corresponding values in the nonrotating limit. As a second step the nonrotating frequencies are expressed as a function of the neutron star average density similar to \cite{Andersson98a}.

\begin{figure}[]
	\centering
	\includegraphics[width=0.55\textwidth]{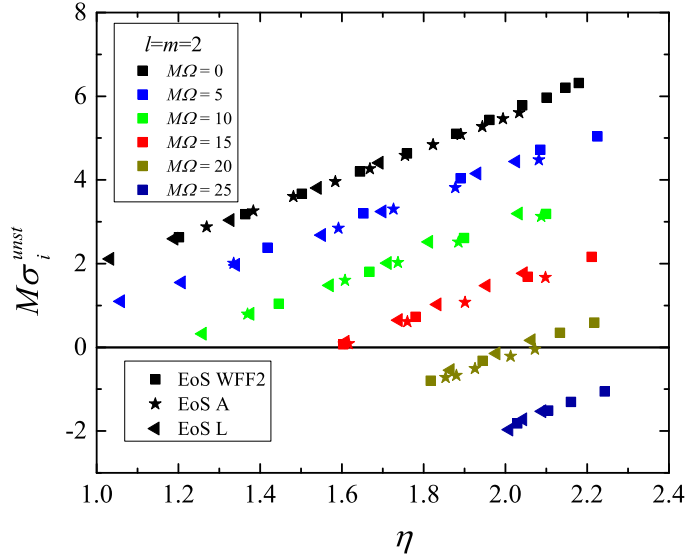}
	\caption{The potentially unstable $l=m=2$ $f$-mode oscillation frequencies measured in inertial frame as functions of the parameter $\eta$.}
	\label{finM_Low_IM_l2}
\end{figure}

Here we follow a different approach. We are going to use the moment of inertia instead of the stellar radius and a different normalization similar to \cite{Lau2010}. It was shown that this makes the asteroseismology relations much more insensitive to the EoS. First, we will focus on counter rotating $f$-modes with $l=m$ that are potentially CFS unstable. In Fig. \ref{finM_Low_IM_l2} we plot the $l=m=2$ $f$-mode oscillation frequencies measured in inertial frame of reference\footnote{The inertial frame frequency $\sigma_i$ is connected to the comoving frame frequency $\sigma_c$ by the relation $\sigma_c = \sigma_i + m\Omega/2\pi$, where $m$ is the azimuthal mode number.} $\sigma_i^{unst}$ normalized to the neutron star mass $M$, as a function of the parameter $\eta$. 
Different colors correspond to sequences with fixed values of the normalized rotational parameter ${\hat \Omega}=M\Omega$ and different styles of the symbols correspond to different EOS. The data range from the nonrotating limit ${\hat \Omega}=0$ to the rapidly rotating case with ${\hat \Omega}=25$ (in units $[ M_\odot\,{\rm kHz}]$). In the latter case the models have rotational frequencies in the range $f=1.2-1.5{\rm kHz}$ and they do not have a stable nonrotating limit. 

The first thing one can notice is that even though the chosen EoS are very different, the data points for a fixed ${\hat \Omega}$ fall very well on a single line. Moreover the slopes of the fitting lines vary by a small amount for different values of ${\hat \Omega}$. It turns out that the dependences can be well approximated by a fitting formula of the type
\begin{equation}\label{eq:Relation_Freq}
{\sigma^{unst}_i M} =(a_1 + a_2 {\hat \Omega} + a_3 {\hat \Omega}^2) + (b_1 + b_2 {\hat \Omega} + b_3 {\hat \Omega}^2) \eta.
\end{equation}
The coefficients $a_1$,..,$b_3$ are given in Table 1 for different values of $l$.

\begin{table}[h] \label{Tbl:Freq_Coeff}\caption{The coefficients $a_1$,..,$b_3$ in the fitting formula \eqref{eq:Relation_Freq}, for the potentially unstable branches with $l=m=2,\,3 \,\,\,{\rm and } \,\,\,4$. }
	\begin{tabular}{|r|rrrrrrr|}
		\hline
		 & $a_1$ & $a_2$ & $a3$ & & $b_1$ & $b_2$ & $b3$ \\
		\hline
		$l=m=2$& $-1.76$ & $-0.143$ & $-6.65 \times 10^{-3}$ &  &  $3.64$ & $-4.36 \times 10^{-2}$ & $2.00 \times 10^{-3}$ \\
		$l=m=3$& $-2.55$ & $-0.236$ & $-1.63 \times 10^{-2}$ &  &  $4.67$ & $-6.63 \times 10^{-2}$ & $5.79 \times 10^{-3}$ \\
		$l=m=4$& $-3.21$ & $-0.365$ & $-2.47 \times 10^{-2}$ &  &  $5.55$ & $-7.33 \times 10^{-2}$ & $9.09 \times 10^{-3}$ \\
		\hline
	\end{tabular}
\end{table}
An interesting fact, we should note, is that the coefficients $a_1$,..,$b_3$ can be expressed as linear functions of the mode number $l$ with a good accuracy (only the error in the fitting of $b_2$ is slightly larger). Thus we obtain
\begin{eqnarray}
{\sigma^{unst}_i M} &=&\left[(-0.332-0.725\, l) + (0.085-0.111\, l) {\hat \Omega} + (0.0112 - 0.00903\, l){\hat \Omega}^2\right] + \notag \\
&&\left[(1.755+0.955\, l) + (-0.0165-0.0149\, l) {\hat \Omega} + (-0.00501 + 0.00355\, l){\hat \Omega}^2\right] \eta.
\end{eqnarray}
An important consequence of this dependence is that with a good accuracy one can say that the relations given by eq. \eqref{eq:Relation_Freq} for different $l$ are linearly dependent. This reflects the possible ways of solving the inverse problem that will be discussed in the next section. 

The above given asteroseismology relations have several advantages compared to the ones considered in \cite{Gaertig10,Doneva2013a}. First, they are much simpler and the rotational frequency $\Omega$ enters at second order compared to the third order relations given in \cite{Gaertig10,Doneva2013a}. Also the deviations from EoS universality are smaller \footnote{As a matter of fact the normalized relations between the oscillation frequencies and the rotational rate in \cite{Gaertig10,Doneva2013a} are also quite independent of the EoS. But the second set of relations, namely between the nonrotating frequencies and the neutron star average density, can lead to larger deviations from the fits as discussed in \cite{Andersson98a,Benhar04}.}. But there are of course also disadvantages compared to the original approach \cite{Gaertig10,Doneva2013a}. Here the asteroseismology relations are prone to the deviations coming from the Cowling approximation, whereas in the two step relations in \cite{Gaertig10,Doneva2013a} it was shown that the normalized relations between the oscillation frequencies and the rotational rate are very close to the results coming from the full nonlinear GR calculations \cite{Zink10,Doneva2013a}.

We will present here also briefly results for the asteroseismology relations for the co-rotating stable modes with $l=-m$. Part of our motivation comes from the fact that by employing a specific normalization of the quantities, we can solve the inverse problem by using three frequencies (both stable and unstable). This was not possible using the relations in \cite{Gaertig10,Doneva2013a} where in order to obtain the mass, radius and rotational rate independently one has to use at least one damping time of a mode.

\begin{figure}[]
	\centering
	\includegraphics[width=0.55\textwidth]{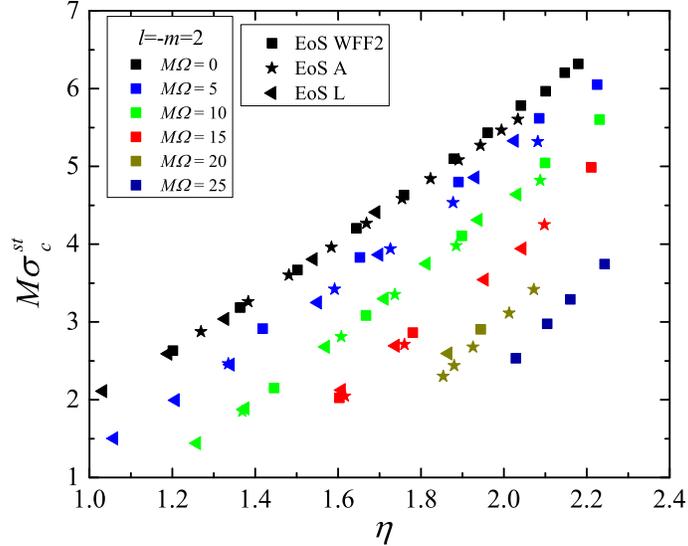}
	\caption{The stable $l=-m=2$ $f$-mode oscillation frequencies in comoving frame as functions of the parameter $\eta$.}
	\label{fig:fcomovM_High_IM_l2}
\end{figure}

The relations for the stable branches are similar to the potentially unstable case. But it turns out that for solving the inverse problem it is better to use the oscillation frequencies in comoving frame $\sigma_c$ instead of $\sigma_i$. Also the dependence on ${\hat \Omega}$ can be well approximated with a linear function compared to the second order polynomial in the potentially unstable case. The normalized dependences between the $l=-m=2$ oscillation frequencies in comoving frame and the parameter $\eta$ are shown in Fig. \ref{fig:fcomovM_High_IM_l2} for different values of ${\hat \Omega}$.

We use a fit very similar to the one for the counter-rotating modes given by eq. \eqref{eq:Relation_Freq} with the only difference that the dependence on ${\hat \Omega}$ is linear: 
\begin{equation}\label{eq:stable_sigma}
{\sigma^{st}_c M} =(a_1 + a_2 {\hat \Omega}) + (b_1 + b_2 {\hat \Omega}) \eta.
\end{equation}
The coefficients $a_1$, $a_2$, $b_1$ and $b_2$ are given in Table II. We present results only for the $l=-m=2$ modes, and not for higher values of $l$, since in general the modes with lower value of $l$ are better emitters of gravitational waves. For the counter-rotating potentially unstable modes we considered also the $l>2$ case because the CFS instability might develop easier for higher $l$ \cite{Gaertig11,Doneva2013a,Passamonti12}. But we do not have such a mechanism for the co-rotating modes and it is generally accepted that only the co-rotating stable $l=2$ modes will emit significant amount of gravitational waves.

\begin{table}[h] \label{Tbl:Freq_Coeff_Stable}\caption{The coefficients $a_1$, $a_2$ and $b$ in eq. \eqref{eq:stable_sigma} for the stable branch with $l=-m=2$. }
	\begin{tabular}{|r|rrrr|}
		\hline
		& $a_1$ & $a_2$ & $b_1$ & $b_2$ \\
		\hline
		$l=-m=2$&  -1.66 & -0.249  & 3.66  & 0.0633 \\
		\hline
	\end{tabular}
\end{table}

\subsection{Damping times}
Deriving asteroseismology relations for the damping time $\tau_{GW}$ is not as straightforward as for the oscillation frequencies for the following reasons. The damping time can be expressed as
\begin{equation}\label{eq:TauGW}
\tau_{GW} =-\frac{1}{2E}\frac{dE}{dt},
\end{equation}
where $E$ is the mode energy and the energy carried by gravitational waves $dE/dt$ is proportional to the squared $l+1$ time derivative of the corresponding multipole moment. Applying the Newtonian multipole formula for the gravitational wave emission one can show that \cite{Balbinski82,Balbinski85,Gaertig10}  
\begin{equation}\label{eq:DampingTime_sigma_multipole}
1/\tau_{GW} \propto \sigma_i^{2l+1} \sigma_c,
\end{equation} 
where $\sigma_i$ and $\sigma_c$ are the frequencies in the inertial and corotating frames respectively. As one can notice $\tau_{GW} \rightarrow \infty$ when $\sigma_i \rightarrow 0$. This means that $\tau_{GW}$ can not be expressed directly as a function of the stellar mass, moment of inertia and rotational rate similar to eq. \eqref{eq:Relation_Freq} because then the peculiar behavior of  $\tau_{GW}$ around the $\sigma_i =0$ point (i.e. around the transition from stable to CFS unstable regime) can not be reflected. Instead $\tau_{GW}$ can be expressed as a function of $\sigma_i$ similar to the previous studies \cite{Gaertig11,Doneva2013a}. 
In order to obtain relations insensitive to the EoS one has to of course use a proper normalization of the quantities, that is explained and justified below.

Making some approximate back of the envelope calculations on the basis of eq. \eqref{eq:TauGW}, and taking into account that the parameter $\eta$ can play the role of an effective compactness instead of $(M/R)$,  one can obtain
\begin{equation}\label{eq:TauRelation_Normalization}
\eta\left(\frac{M}{\tau\eta^2}\right)^{(1/2l)}  \propto M \sigma .
\end{equation}
Even though this formula is based on very simplified assumptions, it offers a good way for normalizing the damping time asteroseismology relations.

\begin{figure}[]
	\centering
	\includegraphics[width=0.45\textwidth]{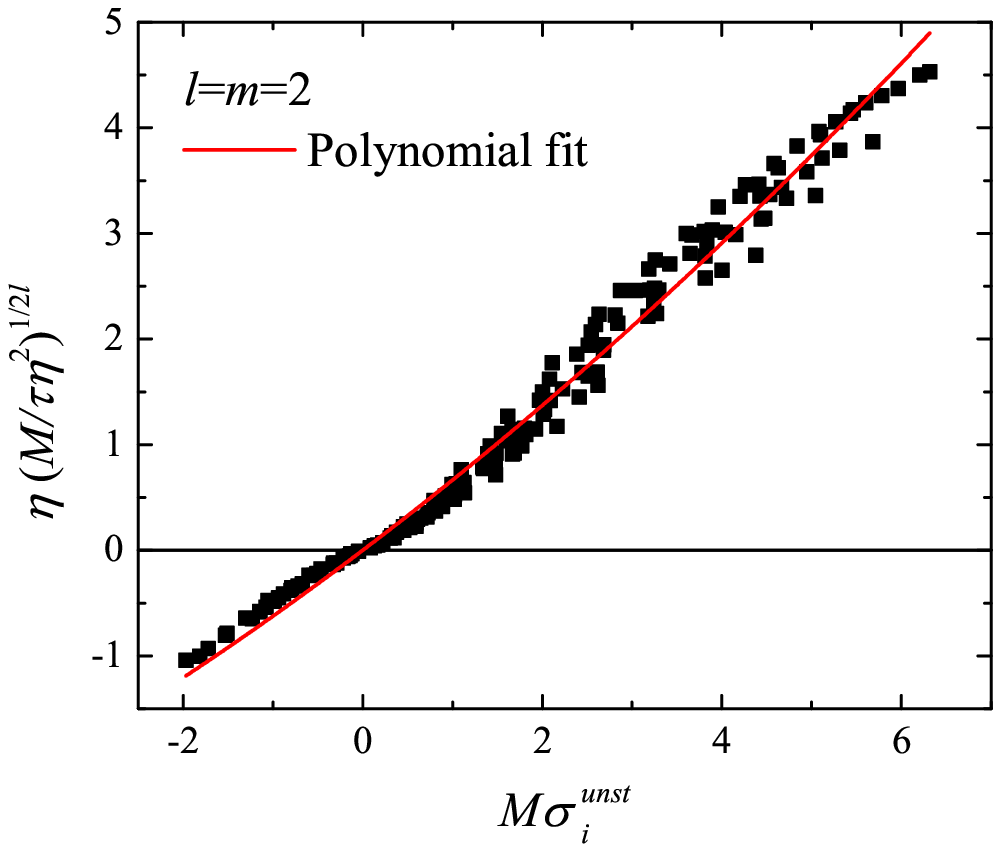}
	\includegraphics[width=0.45\textwidth]{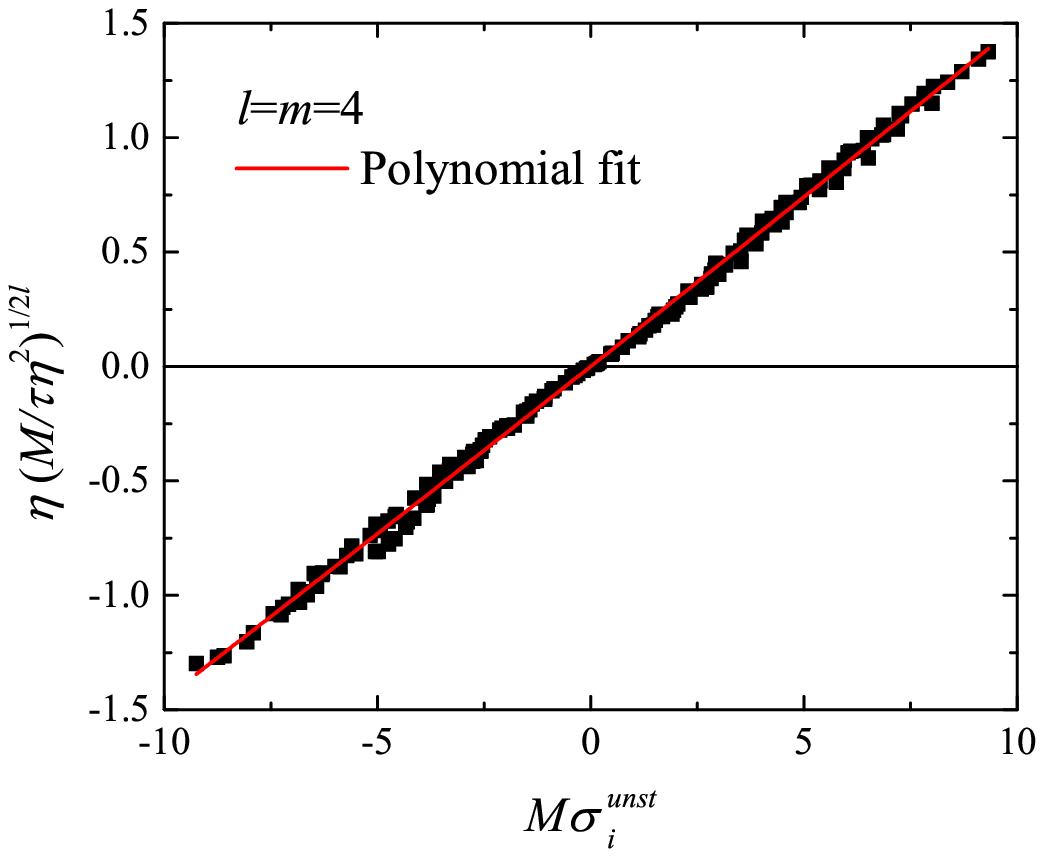}
	\caption{The normalized damping time $\eta\left(M/\tau\eta^2\right)^{(1/2l)}$ as a function of the normalized oscillation frequency $M
		\sigma_i^{unst}$ for the $l=m=2$ and $l=m=4$ $f$-modes.}
	\label{Fig:Asteroseism_Tau}
\end{figure}

The dependences of the normalized  $f$-mode damping time as a function of the normalized oscillation frequency for the potentially CFS unstable branches with $l=m=2$ and $l=m=4$ are shown in Fig. \ref{Fig:Asteroseism_Tau}. One can easily notice that the dependences pass through the origin of the axes where both $\sigma_i^{unst} =0$ and $1/\tau_{GW}=0$ according to eq. \eqref{eq:DampingTime_sigma_multipole}. 
Therefore we can approximate the data with a second order polynomial of the form
\begin{equation}\label{eq:Relation_Damping}
\eta\left(\frac{M}{\tau\eta^2}\right)^{(1/2l)}  =c_1 (M \sigma_i^{unst})+ c_2 (M \sigma_i^{unst})^2,
\end{equation}
where $c_1$ and  $c_2$ are constants. If we substitute $(M \sigma_i^{unst})$ from eq. \eqref{eq:Relation_Freq} into eq. \eqref{eq:Relation_Damping}, we will finally have the desired relation between the damping time from one side and the neutron star mass, moment of inertia and rotational rate from the other.
As one can see from Fig. \ref{Fig:Asteroseism_Tau}, the approximation is very good for $l=m=4$ but the errors in the $l=m=2$ case can be larger. But these large errors are observed mainly in the stable part of the graph (with $\tau_{GW}>0$). That is why we have chosen to give separately the values of $c_1$ and $c_2$ obtained from the fit of the unstable part only. Moreover this is the regime where the CFS instability is operating and can offer a promising mechanism for the emission of detectable gravitational waves signal \cite{Passamonti12,DPK2015}. The values of the coefficients $c_1$ and $c_2$ for $l=m=2,3$ and $4$ are given in Table III both for the full sequence of data and for the CFS unstable part of the branches where $\tau_{GW} < 0$ and $\sigma_i < 0$.

\begin{table}[h] \label{Tbl:Freq_Coeff_Tau}\caption{The coefficients $c_1$ and $c_2$ in eq. \eqref{eq:Relation_Damping} for different values of $l$. }
	Full sequence: \\
	\begin{tabular}{|c|cc|}
		\hline
		& $c_1$ & $c_2$ \\
		\hline
		$l=m=2$& 0.644 & $2.07\times 10^{-2}$ \\
		$l=m=3$& 0.246 & $2.68\times 10^{-3}$ \\
		$l=m=4$& 0.147 & $1.91\times 10^{-4}$ \\
		\hline
	\end{tabular}\\ \mbox{} \\
	Unstable part: \\
	\begin{tabular}{|c|cc|}
		\hline
		& $c_1$ & $c_2$ \\
		\hline
		$l=m=2$& 0.403 & $-7.64\times 10^{-2}$ \\
		$l=m=3$& 0.217 & $-4.82\times 10^{-3}$ \\
		$l=m=4$& 0.144 & $-3.15\times 10^{-4}$ \\
		\hline
	\end{tabular}
\end{table}

Let us now compare the relations derived here with the previous studies. A single dependence between the normalized damping time and the normalized rotational frequency was derived in \cite{Gaertig10,Doneva2013a} for the potentially unstable branches with $l=m=2,3 \,\,{\rm and }\,\, 4$. Thus it was in a way more general than the corresponding relations in the present paper given by eq. \eqref{eq:Relation_Damping}. But as we explained above, the relations in \cite{Gaertig10,Doneva2013a} are two step, and the normalized dependences have to be supplemented with relations between the static damping times and the stellar mass and radius. These static relations are different for different values of $l$ and the deviations from EoS universality can reach large values. That is why if we sum up the advantages and disadvantages in both case, we can conclude that the relations given here are at least as good as the ones in \cite{Gaertig10,Doneva2013a}. Of course in order to determined more precisely which set of relations is better one has to examine also the inverse problem and see how accurately we can derive the stellar properties from the observed mode frequencies and damping times. This will be done in the following section.

We should note that the Cowling approximation can introduce very large deviations in the damping times. For example if the frequency increases with say $30\%$, the damping time decreases by a factor of three (for the $l=2$ mode) due to the strong dependence of $\tau_{GW}$ on $\sigma_i$ according to eq. \eqref{eq:DampingTime_sigma_multipole}. That is why relaxing the Cowling approximation is very important for the exact calculation of the damping times and the related gravitational wave asteroseismology. Such a project is underway.

\section{Solving the inverse problem}
In this section we will consider the inverse problem, i.e. obtaining the stellar parameters using the observed oscillation frequencies and damping times. The previous results for rapid rotation showed that using the normalizations in \cite{Gaertig10,Doneva2013a} one can not obtain the mass, radius and rotational rate of the star independently using only the observed oscillations frequencies due to the specific form of the relations used there. Instead one has to use also the damping times. But observing the damping times is much more difficult compared to the oscillation frequencies and the errors are expected to be considerably bigger. 

The normalization presented in this paper allows us to improve the picture a little bit. Unfortunately one still can not solve the inverse problem using three frequencies of CFS unstable modes with different $l$. The reason is that as we said above $a_1$, .., $b_3$ in eq. \eqref{eq:Relation_Freq} can be expressed with a very good accuracy as linear functions of $l$ which makes the corresponding coefficients for $l=2,3\,{\rm and } \, 4$ linearly dependent. Instead one can supplement the unstable $l=m$ mode frequencies with a frequency of a stable corotating mode having $m<0$. 

In Table IV we present the results after solving the inverse problem in four different potential scenarios:
\begin{enumerate}
	\item We are able to observe the $l=m=2,3$ potentially unstable mode frequencies together with the $l=-m=2$ stable mode frequency.
	\item We are able to observe the $l=m=2,4$ potentially unstable mode frequencies together with the $l=-m=2$ stable mode frequency.
	\item We are able to observe the $l=m=2,3$ potentially unstable mode frequencies together with the $l=m=2$ damping time.
	\item We are able to observe the $l=m=2,4$ potentially unstable mode frequencies together with the $l=m=2$ damping time.
\end{enumerate}
We will represent the accuracy of the inversed asteroseismology relation on two neutron star models that are massive and rapidly rotating and are thus subject to the CFS instability. Such models can be the outcome of core-collapse or binary neutron star merger and they are promising sources of gravitational waves \cite{Passamonti12,DPK2015}. Moreover it was shown that exactly the rapidly rotating massive models are the biggest challenge to the asteroseismology relations presented in  \cite{Gaertig10,Doneva2013a} and lead to large errors when solving the inverse problem. In the cases 3. and 4. where damping time is used as input information, we employ the relations for the unstable part of the asteroseismology relation given in Table III.

\begin{table*}[ht!]
	\begin{center}
		\caption{Solutions of the inverse problem using either three frequencies or two frequencies and a damping time. Two different neutron star models with EoS WFF2 that are subject to the CFS instability are chosen. The rounded percent deviations from the exact values are given in brackets.}\label{tbl:SeismRot_WFF2}
		\begin{tabular}{|c|ccc|cccc|}
			\hline
			& $M [M_\odot]$ & $I/10^{45} {\rm [g/cm^3]}$ & $\Omega [{\rm kHz}]$ &  & $M [M_\odot]$ & $I/10^{45} {\rm [g/cm^3]}$ & $\Omega [{\rm kHz}]$ \\
			\hline
			\bf Exact & 2.02 & 2.80 & 8.66 &  & 2.34 & 3.04 & 9.62 \\[0.1in]
			$\sigma_{l=2}^{unst}$\;\bf\&\; $\sigma_{l=3}^{unst}$\;\bf\&\; $\sigma_{l=2}^{st}$ & 2.02 (0.2) & 2.82 (1) & 8.63 (0.3) &  & 2.39 (2) & 3.17 (4) & 9.48 (1)\\[0.1in]
			$\sigma_{l=2}^{unst}$\;\bf\&\; $\sigma_{l=4}^{unst}$\;\bf\&\; $\sigma_{l=2}^{st}$ & 2.02 (0.2) & 2.82 (1) & 8.63 (0.3) &  & 2.37 (1) & 3.17 (4) & 9.49 (1)\\[0.1in]
			$\sigma_{l=2}^{st}$\;\bf\&\; $\sigma_{l=3}^{unst}$\;\bf\&\; $\tau_{l=2}^{unst}$ & 2.15 (7) & 2.41 (14) & 9.70 (12) &  & 2.35 (1) & 2.72 (10) & 10.12 (5) \\[0.1in]
			$\sigma_{l=2}^{st}$\;\bf\&\; $\sigma_{l=4}^{unst}$\;\bf\&\; $\tau_{l=2}^{unst}$ & 2.17 (7) & 2.41 (14) & 9.75 (13) &  & 2.34 (0) & 2.72 (10) & 10.09 (5) \\[0.1in]
			\hline
		\end{tabular}
	\end{center}
\end{table*}

From the data in Table IV one can make the following conclusions. First, the neutron star parameters can be obtained with a good accuracy using the given combinations of potential observables. This is particularly true for the first two cases with three observed frequencies where the errors are very small. Therefore the relations presented here are more accurate than the ones in \cite{Gaertig10,Doneva2013a} especially in the massive rapidly rotating neutron star case. Also the advantage we have here is that the stellar parameters can be obtained using three frequencies. This is important since the damping times will be much more difficult to be obtained from the detected signal and the observational errors are supposed to be much bigger compared to the oscillation frequencies. Another important point is that the relations presented here are good also for very massive rapidly rotating models that were the biggest challenge to the relations in \cite{Gaertig10,Doneva2013a}. Such neutron stars are of particular interest since they are subject to the CFS instability and are supposed to emit strong gravitational radiation in the early stages of their life \cite{Passamonti12,DPK2015}. That is why obtaining asteroseismology relations that can be used for solving the inverse problem with a good accuracy for such models is important.

It is natural to expect that the accuracy when solving the inverse problem can be improved in two ways. First, as we commented above, is to use a restricted set of EoS, because in this paper we employed a quite broad set in order to test the EoS independence of our relations and some of the EoS are even not in agreement with the current observational constraints. Another way of improving the results is to make a second step in solving the inverse problem explained in detail in \cite{Gaertig10,Doneva2013a}. Namely, we can use the obtained values of $M$, $\Omega$ and $\eta$ and rederive all the asteroseismology relations for values of the parameters close to these particular $M$, $\Omega$ and $\eta$. Afterwards we can use the new relations and obtained updated values of $M$, $\Omega$ and $\eta$. This procedure can be repeated more than once.

To conclude the section let us discuss a possible modification of the asteroseismology relations presented here. A key parameter in the relations is $\eta$ which plays the role of generalized compactness. We used it since it was shown that it leads to the largest known degree of EOS independence   \cite{Lau2010}. But one can obtain similar asteroseismology relations if the true compactness $M/R$ is used instead\footnote{The asteroseismology relations using the compactness $M/R$ were shown to be quite EOS independent in the static limit \cite{Andersson98a,Tsui2005,Staykov2015}.}. The way of building the relations will be practically  the same as the one used in the paper, only $\eta$ has to be substituted with $M/R$. Our rough calculations show that the obtained relations in the case of $M/R$ are also EOS independent up to a large extend. We will not give the explicit relations and numbers here, since the main idea of the paper is not to give as complete set of relations as possible, but instead to demonstrate an alternative way of doing asteroseismology and test its accuracy. We plan to derive a more complete set of relations in the future when we drop the Cowling approximation.

\section{Conclusions}
In the present paper we considered a new way of parameterizing the gravitational wave asteroseismology relations in the case of rapidly rotating neutron stars. We concentrated on the $f$-modes that are one of the most efficient emitters of gravitational radiation. Our study is motivated by the results in \cite{Lau2010} where it was shown that if one normalizes the oscillation frequencies and damping times in a certain way and uses the moment of inertia of the star, the asteroseismology relations become much more insensitive to the nuclear matter equation of state compared to the commonly used relations involving the neutrons star average density \cite{Andersson98a}. The results in \cite{Lau2010} are limited to the static case and here we extended them to rapid rotation and also to values of the spherical mode number $l\ge2$. This is particularly important since the $f$-modes of neutron stars can become CFS unstable for rotational frequencies above roughly 80\% of the Kepler limit \cite{Gaertig11,Doneva2013a,Passamonti12}, thus becoming efficient emitters of gravitational radiation. The case of $l>2$ on the other hand is important since the standard mass neutron stars (with mass say $M<2.0M\odot$) have much larger CFS instability window for modes with $l=m=3$ and $l=m=4$ compared to $l=m=2$.

In our studies we used three different equations of state that cover a very large range a stiffness -- from very soft EOS with small maximum mass, to very stiff EoS with large maximum mass and large radii. Even though some of these EOS are not in the preferred range of neutron star masses and radii according to the observations, they are very useful for testing the EOS independence of our results. It turns our that after choosing an appropriate normalization, the obtained asteroseismology relations are indeed EoS independent up to a large extend for both the oscillation frequencies and damping times.

There are several important differences with the previous results on rapidly rotating neutron stars obtained in \cite{Gaertig10,Doneva2013a}. The main one is that the asteroseismology relations in \cite{Gaertig10,Doneva2013a} are two step. First, we have relations between the normalized oscillations frequency (or damping time) and the normalized rotational frequency, where the normalizations involve the  oscillations frequency and damping times in the nonrotating limit. Second, we have relations connecting  the  oscillations frequency and damping times in the nonrotating limit to the neutron star mass and radius. The normalization considered in the current paper is different -- we have single relation connecting the oscillation frequency (or damping time) to the neutron star parameters (mass, moment of inertia and rotational frequency) for every value of $l$. The relations in the present paper also look simpler -- they are either linear of quadratic with respect to ${\hat \Omega}$ compared to the second and third order relations in  \cite{Gaertig10,Doneva2013a}. Another difference is that we are using the moment of inertia instead of the stellar radius, since it was shown in \cite{Lau2010} that this leads to better EOS independence. On the other hand one can well constrain the nuclear matter EOS using the moment of inertia \cite{Lattimer2005}.

The same methodology as the one described in the present paper can be used to derive asteroseismology relations that employ the stellar radius, as discussed in the previous section. Our rough calculations show that these relations are also quite EOS independent (one just has to substitute the effective compactness $\eta$ with $M/R$). 

At the end we concentrated on different ways to solve the inverse problem, i.e. obtain the stellar parameters from the observed oscillation frequencies and damping times. The most important thing to point out is that using the relations presented in the current paper, we were able to solve the inverse problem with a very good accuracy using three oscillations frequency, that was not possible in the previous studies \cite{Gaertig10,Doneva2013a} where the use of both the mode frequencies and damping/growth times was required. This is quite beneficial, since extracting the damping/growth times from the gravitational wave signal might be very difficult and the accuracy will be much lower compared to the oscillation frequencies. When solving the inverse problem we concentrated mainly on massive models that are rotating rapidly, i.e. the primary candidates for the CFS instability. The derived relations are  particularly good in such cases that represent the biggest challenge to the relations in \cite{Gaertig10,Doneva2013a}, where the errors increase significantly for massive rapidly rotating models.

\section*{Acknowledgements}
DD is grateful to S. Yazadjiev for helpful discussions and suggestions. DD would like to thank the European Social Fund and the Ministry of Science, Research and the Arts Baden-W\"urttemberg for the support. The support by the Bulgarian NSF Grant DFNI T02/6 and "New-CompStar" COST Action MP1304 is gratefully acknowledged.

\bibliography{references}

\end{document}